\documentclass[twocolumn,showpacs,amstex,amssymb,amsmath,prb]{revtex4}
\usepackage{graphicx}
\usepackage{epstopdf}
\begin{document}
\title{Nearsightedness of Electronic Matter in One Dimension}
\author{E. Prodan}
\address{PRISM, Princeton University, Princeton, NJ 08544}
\date{\today }

\begin{abstract}
The  concept of nearsightedeness of electronic matter (NEM) was introduced by W. Kohn in 1996 as the physical principal underlining  Yang's electronic structure alghoritm of divide and conquer. It describes the fact that, for fixed chemical potential, local electronic properties at a point $r$, like the density $n(r)$, depend significantly on the external
potential $v$ only at nearby points. Beyond a distance $\textsf{R}$, changes $\Delta v$ of that potential, {\it no matter how large},  have {\it limited} effects on local electronic properties, which tend to zero as function of $\textsf{R}$. This remains true even if the changes in the external potential completely surrounds the point $r$. NEM can be quantitatively characterized by the nearsightedness range, $\textsf{\textsf{R}}(r,\Delta n)$, defined as the smallest distance from $r$, beyond which {\it any} change of the external potential produces a density change, at $r$, smaller than a given $\Delta n$. The present paper gives a detailed analysis of NEM for periodic metals and insulators in 1D and includes sharp, explicit estimates of the nearsightedness range. Since NEM involves arbitrary changes of the external potential, strong, even qualitative changes can occur in the system, such as the quantization of the energy bands or the filling of the insulating gap of an insulator with continuum spectrum. In spite of such drastic changes, we show that $\Delta v$ has only a limited effect on the density, which can be quantified in terms of simple parameters of the unperturbed system.
\end{abstract}

\pacs{71.10.-w, 71.15.-m}

\maketitle

\section{Introduction}

This paper is based on a preliminary remark by W. Kohn,\cite{Kohn96} about a general concept called ``nearsightedness of electronic matter (NEM)" and on a recent short report (PK),\cite{ProdanKohn} which amplified that remark in various aspects but did not include detailed proofs. In the present paper,
we select that part of PK dealing with non-interacting 1D electrons and provide a full discussion, including detailed proofs. Future publications will amplify other sections of PK.

By ``electronic matter" we understand a system of many electrons with significant wavefunction overlap, in equilibrium under the action of a given external potential $v(r)$. We shall consider the change of a local electronic property, like the electron density $n(r)$, under the action of an arbitrarily strong
potential perturbation $w(r')$, which vanishes inside a specified sphere, $|r-r'|=\textsf{R}$. Note that we allow situations when the perturbation completely surrounds the point $r$. NEM states that the resulting density change at $r$, $\Delta n(r,\textsf{R})$, is bounded by a function $\overline{\Delta n}(r,\textsf{R})$,
\begin{equation}\label{NEM1}
 \Delta n(r,\textsf{R})\leq \overline{\Delta n}(r,\textsf{R}),
\end{equation}
independent of the amplitude or shape of $w(r')$, and that
\begin{equation}\label{NEM2}
\overline{\Delta n}(r,\textsf{R})\rightarrow 0, \
\text{monotonically as}\ \textsf{R}\rightarrow \infty,
\end{equation}
provided only that the chemical potential $\mu$ is held fixed.

The essence of NEM is contained in Eq.~(\ref{NEM1}). Although it may not look very special, consider our perturbing potential $w(r')$, confined outside the sphere of radius $R$. The common sense says that, as we increase its strength, we need to increase the radius $R$ if we want to maintain its effect at the center of the sphere below a certain level. The common sense also says that we need to increase $R$ to infinity as we make $w(r')$ stronger and stronger. In reality, this is not so: if the chemical potential is kept fixed, these effects will saturate and, in fact, no matter what $w(r')$ we put outside the sphere of radius $R$, they cannot exceed a certain upper bound, which we will determine in this paper.  

For a given $\Delta n$, we can solve for $\textsf{R}$ in $\overline{\Delta n}(r,\textsf{R})=\Delta n$ and define the nearsightedness range $\textsf{\textsf{R}}(r,\Delta n)$. The significance of $\textsf{R}(r,\Delta n)$ is the following: any perturbation beyond $\textsf{R}(r,\Delta n)$, of arbitrary shape and
amplitude, cannot produce a density change at $r$ larger than $\Delta n$. $\textsf{R}(r,\Delta n)$ provides a simple, quantitative measure of nearsightedness.

The above formulation of NEM often reminds people of Thomas-Fermi screening, sometimes even when we discuss insulators. However, let us point a few facts. If one puts a charge inside the uniform charged electron gas and calculates the density response, he will find that the Thomas-Fermi exponential screening is valid only very near the impurity. Further away from the impurity, he will see Friedel oscillations, decaying as an inverse power law.\cite{Friedel} These oscillations are not negligible. In fact, the Friedel oscillations were observed experimentally, thought not directly, one year after their theoretical prediction.\cite{Rowland1} However, this was not realized until much later, when Walter Kohn made the connection between the two results.\cite{Rowland2, Kohn60} He showed that there is a big discrepancy between the prediction of the Thomas-Fermi screening theory and these experimental results and that a self-consistent calculation along Friedel's lines, of the density change due to impurities in coper, brings the theory and experiment to almost perfect agreement. The picture that emerged was that, in the asymptotic region, the screening only renormalizes the amplitude of the Friedel oscillations. The whole issue was considered very important at that time, because it clearly demonstrated the existence of a sharp Fermi surface in real metals. We also like to mention one very well known fact in surface physics, where self-consistent calculations of metallic surfaces showed that the effective potential goes to the bulk value extremely fast, typically within one or two layers. However, the density oscillations extend much further into the bulk and they can be viewed as the Friedel oscillations generated by the screened surface potential. With these being said, we hope that the reader will dissociate, right from the begenning, NEM from the Thomas-Fermi exponential screening and the nearsightedness range from the Thomas-Fermi screening length.

Quantum gases display non-local density responses to local perturbations because of two factors. First, the effect of any local perturbation propagates to further distances through inter-particle interactions. This effect can be regarded as classical, since it manifests, in the same way, in classical gasses. Secondly, there is a purely quantum effect, that steams directly from the uncertainity principle. This paper is concerned with this purely quantum effect, so it neglegts the inter-particle interation effects, entirely. The fact that NEM exists for non-interacting systems is extremely important. To understand why, let us go back in time and recall that, at the beginning of the electronic structure calculations, when the exact diagonalization was the method of choice, people were facing the so called ``exponential wall'' when trying to extend the calculations to larger systems: because the number of operations in such calculations scales exponentially with the number of atoms, $N$, their applicability was, and is still limited to systems containg a few tens of atoms. Density Functional Theory (DFT)\cite{KohnHohenberg,KohnSham} provided a powerful alternative: because the number of operations in DFT calculations scales as $N^3$, we can now solve the electronic structure for systems containing hundreds of atoms. However, electronic structure calculations for biological and nano systems, or for extremely complex materials, involve thousands of atoms. At this scale, we start feeling the ``$N^3$ wall.'' {\it Ab-initio} quantum calculation for such complex systems will require a new generation of DFT algorithms, scaling linearly with the number of atoms. It is now generally accepted that NEM is the physical basis for these algorithms.\cite{Giullia}

The quest for linearly scaling algorithms was initiated by W.T. Yang, who was the first to argue that O(N) algorithms are possible.\cite{Yang91} The algorithm proposed by Yang is known by the name of Divide and Conquer (DC). There are now several reviews on the linear scaling electronic structure calculations. We will mention here the one by Goedecker\cite{Goedecker} and the one by Wu and Jayanthi,\cite{Wu} which, at the time of their publication, gave an exhaustive discussion of O(N) methodologies. If we examine carefully these methodologies, they are all based on the same original idea, namely gluing together calculations done for smaller systems. What is different, is the representation and the way the size of these smaller systems (the truncation) is determined. For example, the real space approaches will use the decay of the density matrix while the localized basis set approaches will use the overlap of these functions to judge how large these subsystems should be.

Let us focus on the original implementation given by Yang.\cite{Yang91} Consider a self-consistent DFT iteration process, for a large quantum system. Each iterative step consists in calculating the density of a non-interacting electron gas in equilibrium under a given effective potential (known from the previous iteration). In traditional approaches, this requires a number of operations that scales as $N^3$. DC algorithm, if it works as it is supposed to, requires a number of operations that scales linearly with $N$. It goes like this: The large system is devided in non-overlaping sub-regiones, which are then surrounded with buffer zones. A global chemical potential, $\mu$, is fixed and the orbitals are calculated and populated up to $\mu$, for each sub-region + buffer zone. The density in the buffer zones is discarded so, at this point, one has calculated a density for each sub-region and, by putting together all these sub-densities, one can construct the global density. The charge neutrality condition is then checked for this global density and the chemical potential is adjusted, if necessary. Note that charge neutrality must be satisfied by the entire system, not by each subsystem. In this way we have completed the DFT iteration step in a number of operations that scales linearly with the size of the system. Now, the question is how accurate is this algorithm? In fact, the most important question is, can we obtain arbitrary accuracy with this algorithm? To answer, we need to compare the density calculated for a sub-region + buffer zone and the density calculated for the entire system at once, at the {\it same} chemical potential $\mu$. Now one can see why NEM can be regarded as the basis for this algorithm: the artificial termination at the outer boundary of the buffer zone, no matter how it is done, represents the change in the effective potential in our formulation of NEM. For example, such changes of the effective potential occure when one calculates the density matrix and ignores the points outside a sub-region, or when one calculates the density and ignores the elements of a localized basis set that are centered outside a sub-region. Since we have no control on how the effective potential is modified by such truncations, Eq.~(\ref{NEM1}) is paramount: it tells us that the effects of any artificial termination cannot exceed an upper bound. This upper bound is an intrinsic characteristic of the system: is independent of the method of termination. Eq.~(\ref{NEM2}) tells us that, if we take the buffer zones large enough, the difference between the sub-density and the real density can be made smaller than any desired accuracy. 

Examples and estimates of the width of the buffer zones, together with a discussion of how to optimize this algorithm in 1, 2 and 3D and how the CPU time scales with the the desired accuracy can be found in Ref.~\onlinecite{ProdanKohn}. We also like to mention that DC has been recently implemented to systems containing as many as 65,000 atoms and shown that it can generate electronic structures of the same quality as a traditional approach will do for a system of, let us say, 10 atoms.\cite{Vashishta05} The tests performed in this numerical work agree qualitatively with our theoretical predictions. 

There is another important issue related to DC. The ground energy in DFT is given by:
\begin{equation}
E=\sum_j \epsilon_j+E_{xc}[n]-\int v_{xc}(r)n(r)-\frac{1}{2}\int \frac{n(r)n(r')}{|r-r'|}.
\end{equation}
All the above terms can be calculated directly from the density, except the first one. However, this term is just the integral of the energy density,
\begin{equation}
    \epsilon(r)=\sum_{\epsilon_i\leq \epsilon_F}\epsilon_i
    |\psi_i(r)|^2,
\end{equation}
and we will show that $\epsilon(r)$ is also nearsighted. As a consequence, within the DC algorithm, $\epsilon(r)$ can be calculated with arbitrary precision, like the density.

The goal of this paper is two-fold. We want to prove  NEM, i.e. Eqs.~(\ref{NEM1}) and (\ref{NEM2}), for a simple system, which is the 1D non-interacting, spin $\frac{1}{2}$ fermions, in periodic potentials, and also want to show that one can obtain exact estimates of the nearsightedness range, which is extremely important for DC. The case of 1D non-interacting electrons is important for several reasons. Inspite of its simplicity, it captures all the important aspects of nearsightedness. This allows for a thorough investigation of NEM, while keeping the technical aspects at a reasonable level. The 1D non-interacting case is relevant for linear molecular chains when treated within DFT.

\section{The strategy}

We shall first develop general tools that will allow us to compute, for arbitrary perturbations, the asymptotic behavior of the density change,
\begin{equation}
	\Delta n(x)=2\sum_{\epsilon_i \leq \mu}|\psi_i(x)|^2-2\sum_{\epsilon_i^0 \leq \mu}|\psi_i^0(x)|	^2,
\end{equation}
where $\psi_i^0(x), \epsilon_i^0$ and $\psi_i(x), \epsilon_i$ are the wave functions and the corresponding energies of the unperturbed and perturbed systems, respectively. The factor 2 in front of the sums comes from the spin. The above expression is not very useful when dealing with the asymptotic behavior of $\Delta n(x)$. Instead, we will work with an integral representation. Why an integral representation? For answer, we point to the theory of special functions, where the functions are most often defined and introduced as infinite series but, with no exception, their asymptotic behavior is derived from equivalent integral representations. 

We can obtain an integral representation of $\Delta n(x)$ by using the Green's functions. Indeed, if $G_E^0\equiv (E-H_0)^{-1}$ and $G_E\equiv (E-H)^{-1}$ denote the Green's functions of the unperturbed and perturbed systems, respectively, then
\begin{equation}\label{IntG}
	\Delta n(x)=\frac{1}{\pi} \int_{\cal C} [G_E-G_E^0](x,x) \ dE,
\end{equation}	
where ${\cal C}$ is a contour in the complex energy plane, surrounding the occupied states.
This can be seen from the eigenfunction expansions of $G_E^0$ and $G_E$ and the residue theorem. Now, the eigenfunction expansions of  $G_E^0$ and $G_E$ are, again, not very useful. Instead, we will use the following compact representation:
\begin{equation}\label{Sturm}
	G_E(x,x^\prime)=\frac{\psi_<(x_<)\psi_>(x_>)}{W(\psi_<,\psi_>)},
\end{equation}
where $x_<=\min(x,x^\prime)$ and $x_>=\max(x,x^\prime)$; $\psi_<(x)$ and $\psi_>(x)$ are the solutions of the Schrodinger equation at energy $E$, satisfying the boundary condition to the left and right, respectively, and $W(\psi_<,\psi_>)$ is the Wronskian of the two solutions. For infinite systems, the case considered in this paper,  $\psi_<(x)$ and  $\psi_>(x)$ are the solutions decaying at $\pm \infty$, respectively. We will always assume that $E$ does not belong to the energy spectrum. When the contour ${\cal C}$ intersects the energy spectrum, such as for the case of metals, $G_E$ will have a discontinuity at the point of intersection. Strictly speaking, this point must be excluded from ${\cal C}$, which does not change the result of the integration. For all the other points of ${\cal C}$, $G_E$ is uniquely defined and given by Eq.~(\ref{Sturm}). Later, we will use the reflection and transmission coefficients to construct extremely simple and compact expressions of the Green's functions (see Eqs.~(\ref{GreenF}), (\ref{Fundamental}) and (\ref{DiffG})).

The last step of our strategy will be to identify the special point in the complex energy plane that determines the asymptotic behavior of the integral Eq.~(\ref{IntG}).

This strategy will require from us to go into the complex energy plane. We will make use of the analytic structure of the Bloch functions and band energies derived in Ref.~\onlinecite{Kohn59}, which is briefly discussed in the next section. These analytic structure results can be generalized to linear molecular chains and even to 3D crystals.\cite{Prodan05} Also, the above expression for the Green's function, Eq.~(\ref{Sturm}), can be generalized to linear molecular chains,\cite{Prodan05} or to 3D crystals.\cite{Allen} In fact, the entire strategy can be applied in 2 and 3D, as it was already shown in Ref.~\onlinecite{ProdanKohn}.

\section{The unperturbed system}

Throughout this paper, $v(x)$ will be taken as a periodic, inversion-symmetric potential. Following is a brief discussion, largely taken from Ref.~\onlinecite{Kohn59}, of the periodic Schrodinger equation.

\begin{figure}
  \includegraphics[width=8.6cm]{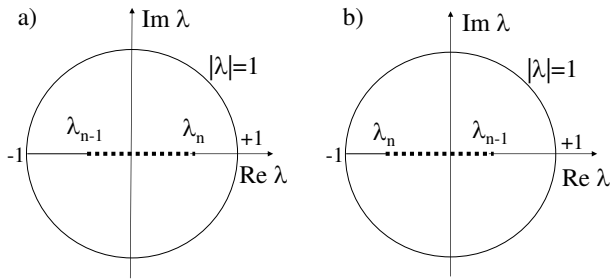}\\
  \caption{The analytic domain of $E_{n}(\lambda)$ and $\psi _{n,\lambda }(x)$  consists of the unit disk except a branch cut, represented by the dotted line.  Panel a) refers to $n$  even and panel b) to $n$ odd.}
\end{figure}

The solutions of the periodic Schrodinger equation,
\begin{equation}\label{Schrodinger}
    [-d^{2}/dx^2+v(x)] \psi =E\psi , \ \ v(x+b)=v(x),
\end{equation}
are the well known Bloch functions $\psi _{k}$ ($k$ the wave vector) which will be normalized as in Ref.~\onlinecite{Kohn59}. Their fundamental property is $\psi_k(x+b)=e^{ikb}\psi_k(x)$. When dealing with complex values of $k$, is much more convenient to work with the variable $\lambda=e^{ikb}$, instead of $k$. Thus, from now on, we will index the Bloch functions by $\lambda$; their fundamental property becomes:
\begin{equation}\label{Bloch}
\psi_{\lambda}(x+b)=\lambda\psi_{\lambda}(x).
\end{equation}
The parameter $\lambda$ relates to the energy of $\psi_{\lambda}$ through the following equation:
\begin{equation}\label{lambdaEq}
    \lambda ^{2}-2\mu(E)\lambda +1=0,
\end{equation}
whith $\mu(E)$ the Kramers' function.\cite {Kramers35} 
By examining the fundamental property Eq.~(\ref{Bloch}), one can see that the physical states correspond to the case $|\lambda|=1$ ($\lambda$ on the unit circle), otherwise $\psi_\lambda(x)$ explodes either to $x=\infty$ or $x=-\infty$. For $\lambda$ on the unit circle, the solutions behave like waves, thus it is appropriate to use the term Bloch waves. When discussing arbitrary values of $\lambda$, however, it is more appropriate to use the term Bloch functions.

The energy spectrum consists of energy bands, indexed here by $n=1,2,\ldots$, which are separated by energy gaps. The energy bands can be computed by solving for $E$ in Eq.~(\ref{lambdaEq}) for all $\lambda$ on the unit circle. Due to the symmetry $\lambda\rightarrow 1/\lambda$ in Eq.~(\ref{lambdaEq}), we can and shall restrict $\lambda$ to $|\lambda | \leq 1$, and view $\psi_\lambda (x)$ and $\psi_{1/\lambda} (x)$ as two independent wavefunctions. For $|\lambda|<1$, it follows from the fundamental property, Eq.~(\ref{Bloch}), that $\psi_\lambda(x)$ decays to zero as $x\rightarrow \infty$ and $\psi_{1/\lambda}(x)$ decays to zero as $x\rightarrow -\infty$.

If we restrict $\lambda$ to the unit disk, $E$ uniquely determines $\lambda $. The opposite is not true, instead $E(\lambda)$ is a multi-valued complex function, with branch points of order one at $\lambda _1$, $\lambda _2$, \ldots. Each $\lambda _n$ is real and $0<(-1)^n\lambda_n<1$. The corresponding energy, $\tilde{E}_n\equiv E(\lambda_n)$, is also real and located in the $n$-th gap. Estimates of $\lambda_n$ in the small gap and tight binding limits are given in Appendix A. $E(\lambda)$ can be represented on a Riemann surface with one sheet corresponding to one band. The $n$-th sheet can be taken as the entire unit disk, except a cut extending from $\lambda _{n-1}$ to $\lambda_n$, as shown in Fig.~1. Near a branch point, $E(\lambda)$ behaves as the square root, 
\begin{equation}\label{Eexpansion}
    E(\lambda)=\tilde{E}_n+2\alpha_n(\lambda/\lambda_n-1)^{1/2}+\ldots.
\end{equation}
The function $E(\lambda)$ on the $n$-th Riemann sheet will be denoted by $E_n(\lambda)$. 

\begin{figure}
  \includegraphics[width=8.6cm]{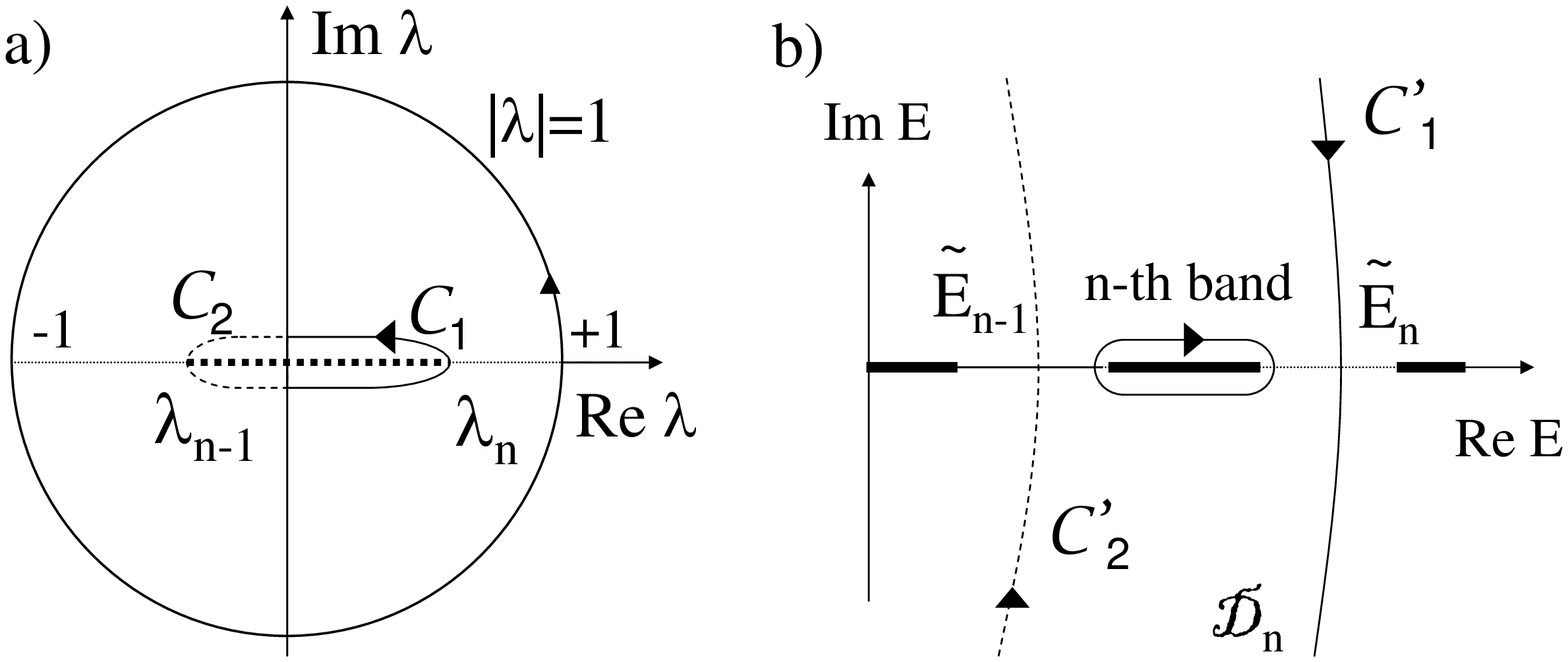}
  \caption{Illustration of how $E_{n}(\lambda)$ maps a)  the complex $\lambda$-plane into b) the complex $E$-plane ($n$ even).}
\end{figure}

The integral in Eq.~(\ref{IntG}) will be mapped into the complex $\lambda$-plane, by changing the variable from $E$ to $\lambda$. Thus, it will be important to understand how the contour ${\cal C}$ looks in this plane. It is more easy to understand how a given contour $\gamma$ in the $\lambda$-plane  is mapped in the complex energy plane, i.e. to construct the points $E(\lambda)$ when $\lambda$ sweeps through the points of $\gamma$. The mapping $E_n(\lambda)$, from the unit disk to the complex $E$-plane, is generic in one dimension, in the sense that it is qualitatively independent of the periodic potential. The general picture is as follows:  $E_n(\lambda)$ maps the unit disk into a domain ${\cal D}_n$ (see Fig.~2). The domains ${\cal D}_n$, $n=1,2,\ldots$, are disjoint (with the exception of a possible common boundary), and, all together, they cover the entire complex $E$-plane. Now let us consider several contours. In Fig.~2, we used the same line-style for a contour and its image. The thick lines in Fig.~2b represent the energy bands. Now consider the contours ${\cal C}_{1}$
  and ${\cal C}_{2}$ in the $\lambda$-plane, starting and ending at zero and surrounding the
  branch cuts infinitely close. They are mapped into ${\cal C}_1^\prime$ and
  ${\cal C}_2^\prime$, shown in Fig. 2b. This figure displays only a finite sector of the two contours, which extend from $-i\infty$ to $+i\infty$. The unit circle is mapped in a loop that surrounds the $n$-th band, infinitely close. The segment from $-1$ to $\lambda _{n-1}$ is mapped on the real axis, from the lower  edge of the $n$-th band down to $\tilde{E}_{n-1}$. The segment from $\lambda _n$ to $+1 $ is also mapped on the real axis, from $\tilde{E}_n$ down to the upper edge of the $n$-th band. The domain ${\cal D}_n$ mentioned above, lies between the curves ${\cal C}_1^\prime$ and ${\cal
  C}_2^\prime$. From the above information, one should be able to construct, qualitatively, the image of any other contour.

$\psi_{\lambda}(x)$ and $\psi_{1/\lambda}(x)$ are multi-valued analytic functions of $\lambda $, with branch points of order 3 at $\lambda_n$. For $\lambda$ on the $n$-th Riemann sheet, $\psi_{\lambda}(x)$ will be denoted by $\psi_{n,\lambda}(x)$. Both functions diverge at the branch points as,
\begin{equation}\label{psi}
    \psi_{n,\lambda}(x)=\frac{u_n(x)e^{-q_n x}}{(\lambda /\lambda
    _{n}-1)^{1/4}}+\ldots
\end{equation}
and
\begin{equation}\label{psip}
    \psi_{n,1/\lambda}(x) =\frac{u_{n}^{\prime }(x)e^{q_n x}}{(\lambda
    /\lambda_{n}-1)^{1/4}}+\ldots,
\end{equation}
where $u_n(x)$ and $u_n^{\prime}(x)$ are periodic (antiperiodic) functions for $n$ even (odd) and $q_n$ is defined by
$|\lambda_n|=e^{-q_n b}$. The Wronskian of the two independent Bloch functions is given by
\begin{equation}
    W(\psi_{n,1/\lambda},\psi_{n,\lambda }) =-\frac{b\lambda }{2\pi }
    \frac{dE_n(\lambda)}{d\lambda}.
\end{equation}
Consequently, the Green's function $G_E^0$ satisfies
the identity:
\begin{equation}\label{GreenF}
    G_E^0(x,x^\prime) \frac{dE}{\pi i}=2i\psi_{n,1/\lambda}(x_<)
    \psi_{n,\lambda}(x_>)\frac{d\lambda}{b \lambda}.
\end{equation}

\section{The Effect of Perturbations}

\subsection{One Sided Perturbations}

We consider here perturbations that are either to the left or to the right of the point $x$, where we measure the density change $\Delta n(x)$. Let us assume that $w$ is to the left of $x$. For convenience, we choose the origin of $x$ at the right edge of $w$, so that $w$ is confined in the interval $[-L,0]$, with $L$ arbitrarily large but finite. We calculate the particle and energy density changes at $x>0$.

As already mentioned, the density change $\Delta n(x)$ is given by 
\begin{equation}\label{prima}
    \Delta n(x)=\frac{1}{\pi i}\int_{\cal C} [G_E(x,x)-G_E^0(x,x)]dE,
\end{equation}
where $G_E^0$ and $G_E$ are the unperturbed and perturbed Green's functions, respectively, and ${\cal C}$ is a contour in the complex energy plane, surrounding the eigenvalues below $\epsilon_F$ (see for example Fig.~3a). Similarly, the change of the energy density is
\begin{equation}\label{duo}
    \Delta \epsilon(x)=\frac{1}{\pi i}\int_{\cal C}
    E[G_E(x,x)-G_E^0(x,x)]dE.
\end{equation}
We can focus on $\Delta n(x)$ and give only the final results for
$\Delta \epsilon (x)$.

We construct the perturbed Green's function from two independent
solutions of the Schrodinger equation:
\begin{equation}
    [-d^2/dx^2+v(x) +w(x)]\psi(x)=E\psi(x).
\end{equation}
 Outside the interval $[-L,0]$, the solutions are linear combinations of $\psi_{\lambda}$ and $\psi_{1/\lambda}$. As already mentioned, we need the solutions decaying to $\mp\infty $, which can be conveniently written in terms of the reflection and transmission coefficients:
\begin{equation}\label{psiless}
    \psi _{n,\lambda }^{<}(x) =\left\{
    \begin{array}{l}
       T_n(\lambda) \psi_{n,1/\lambda}(x), \ x<-L \\
      \psi_{n,1/\lambda}(x)+R_n^{+}(\lambda) \psi_{n,\lambda}(x),
      \ x>0
\end{array}
\right.
\end{equation}
and
\begin{equation}\label{psimore}
    \psi_{n,\lambda }^{>}(x) =\left\{
    \begin{array}{l}
        \psi_{n,\lambda }(x)+R_n^- (\lambda)\psi
        _{n,1/\lambda}(x), \ x<-L \\
        T_n(\lambda)\psi_{n,\lambda}(x), \ x>0,
    \end{array}
\right.
\end{equation}
where $E$ was taken to be in ${\cal D}_{n}$ (Fig.~2). The Wronskian of the two independent solutions is
\begin{equation}
    W(\psi_{n,\lambda}^{<},\psi_{n,\lambda }^{>}) =-\frac{b}{2\pi
    }\lambda T_{n}(\lambda) \frac{dE_{n}(\lambda) }{d\lambda },
    \label{Wronskian}
\end{equation}
leading to the following useful identity, for $x>0$:
\begin{equation}\label{Fundamental}
    [G_E-G_E^{0}](x,x)\frac{dE}{\pi i}
    =2iR_{n}^{+}(\lambda)\psi_{n,\lambda}(x)^2
    \frac{d\lambda }{b\lambda }.
\end{equation}
This identity, together with Eqs.~(\ref{prima}) and (\ref{duo}), shows that, for $x>0$, $\Delta n(x)$ and $\Delta \epsilon (x)$ are completely determined by the unperturbed wavefunctions and reflection coefficient, in a simple and universal way.

\begin{figure}
  \includegraphics[width=8.6cm]{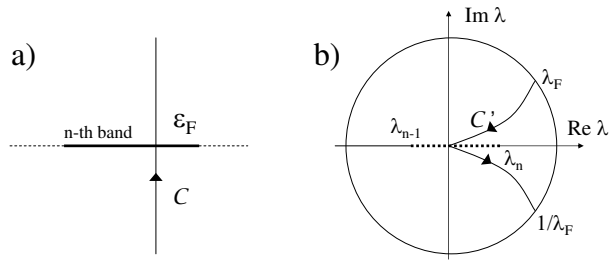}\\
  \caption{The contour of integration for metals in a) complex $E$-plane
  and b) complex $\lambda$-plane. The contour in panel a) extends from $-i\infty$ to $+i\infty$.}
\end{figure}

Eq.~(\ref{Fundamental}) also provides the analytic structure of the reflection coefficient: $R_{n}^+(\lambda )$ has branch points of order 1 at $\lambda_{n-1}$ and $\lambda_{n}$. If we go around these
branch points, $R_{n}^+$ becomes $R_{n-1}^{+}$ and $R_{n+1}^+$, respectively. In other words, $R_{n}^+$ are different branches of a multi-valued function $R^+(\lambda)$. Near the branch points,
\begin{equation}
    R^+(\lambda)=R^+(\lambda_{n})+r_n^+(\lambda/\lambda_n-1)^{1/2}+\ldots.
\end{equation}
The poles of $R^{+}(\lambda)$, if any, are mapped by $E(\lambda)$ into the poles of $G_E$, i.e. the energies of the bound states. Since the bound states are located in the gaps, the poles of $R^{+}(\lambda)$ are always located on the real axis and away from the branch cuts.

By evaluating the Wronskian of $\psi_{\lambda}^<(x)$ and $\psi_{1/\lambda}^<(x)$ for $x<-L$ and for $x>0$ and equating the two results, one can derive the following identity:
\begin{equation}
    T(\lambda) T(1/\lambda)+R^{+}(\lambda)R^{+}(1/\lambda) =1.
\end{equation}
For $|\lambda| =1$, this identity becomes $|T(\lambda)|^2+|R^{+}(\lambda)|^2=1$, showing that
$|R^{+}(\lambda)|\leq 1$ for all $\lambda$ on the unit circle. Similar conclusion holds for $R^{-}(\lambda)$.

\subsubsection{Metals}

\begin{figure}
  \includegraphics[width=7.6cm]{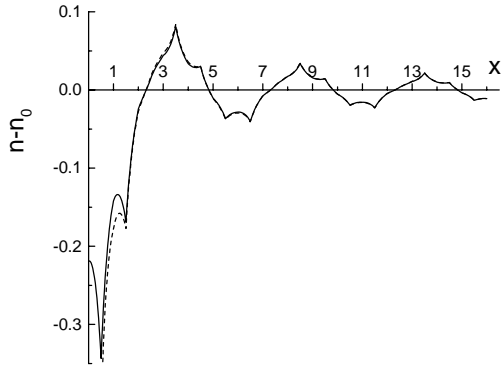}\\
  \caption{The exact (solid) and asymptotic (dashed)
  $\Delta n(x)$ for the model of Eq.~(\ref{vtotal})
  [$v_{0}=-2$, $V_{0}=10$, $b=1$, first band 20\% filled].}
\end{figure}

We assume the $n$-th band partially occupied and choose ${\cal C}$ in Eq.~(\ref{prima}) as in Fig.~3, where $\lambda_F$ is defined by $\epsilon_F=E_n(\lambda_F)$. We write 
\begin{equation}\label{m}
    x=y+mb,
\end{equation}
with $y$ restricted to the first unit cell. Mapping into the $\lambda$-plane by using Eq.~(\ref{Fundamental}) and recalling the fundamental property of the Bloch functions, for $x>0$, we obtain
\begin{equation}
    \Delta n(x) =\frac{2i}{b}\int_{C^{\prime}}R_{n}^{+}(\lambda)\psi_{n,\lambda }(
    y)^2\lambda^{2m-1}d\lambda.
\end{equation}
An integration by parts gives
\begin{eqnarray}
    \Delta n(x)&=&2\text{Im}\frac{R_n^+(\lambda_F)\psi_{n,\lambda_F}(x)^2}{mb}  \nonumber \\
    &-&\frac{i}{b}\int_{C^\prime}\frac{\lambda^{2m}}{2m}\frac{d}{d\lambda }\left[
    R_n^+(\lambda)\psi_{n,\lambda}(y)^2\right]d\lambda.
\end{eqnarray}
The integral is of order $1/x^2$, as it can be seen from another integration by parts. We can conclude
\begin{equation}\label{asymptotic3}
    \Delta n(x) \rightarrow \frac{2}{x}\text{Im}[R_n^+(\lambda
    _F)\psi_{n,\lambda_F}(x)^2],
\end{equation}
for large $x$. Similarly,
\begin{equation}
    \Delta \epsilon(x) \rightarrow \frac{2\epsilon_F}{x}\text{Im}[R_n^+(\lambda
    _F)\psi_{n,\lambda_F}(x)^2].
\end{equation}
Since $|R_{n}^{+}(\lambda_{F})|\leq 1$, the amplitudes of $\Delta n(x)$ and $\Delta \epsilon(x)$ cannot exceed, in the asymptotic
limit, the upper bounds
\begin{equation}\label{upper1}
    \overline{\Delta n}(x) \rightarrow
    \frac{2}{x}|\psi_{n,\lambda_{F}}(x)|^2,
\end{equation}
and
\begin{equation}
    \overline{\Delta \epsilon}(x) \rightarrow
    \epsilon_F \overline{\Delta n}(x),
\end{equation}
independent of the shape and amplitude of $w(x)$. This proves NEM for metals and one sided perturbations.

A comparison between Eq.~(\ref{asymptotic3}) and an exact calculation of $\Delta n(x)$ for the perturbed Kronig-Penney
model\cite{Kronig}
\begin{equation}
    v_{t}(x) =\left\{
    \begin{array}{l}
        v_{0}\sum_{l=-\infty }^{\infty }\delta \left( x+\frac{b}{2}-lb\right) ,\,x>0\\
        V_{0},\,x\leqslant 0,
    \end{array}
    \label{vtotal}
    \right.
\end{equation}
is shown in Fig.~4. Note that the asymptotic regime starts from about two lattice constants away from the perturbation.

\subsubsection{Insulators}

\begin{figure}
  \includegraphics[width=8.0cm]{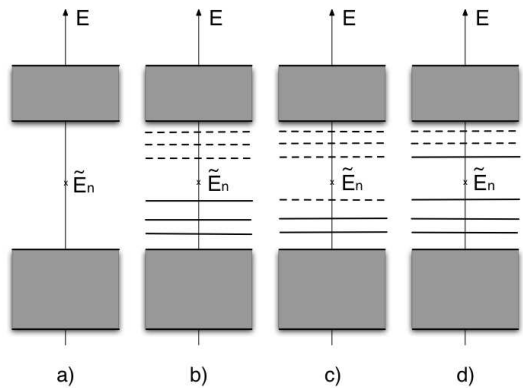}\\
  \caption{a) No bound states are present in the insulating gap. b) Bound states are present, but all the states below $\tilde{E}_n$ are occupied (solid lines) and all the states above $\tilde{E}_n$ are unoccupied (dashed lines). c) States below $\tilde{E}_n$ are unoccupied. d) States above $\tilde{E}_n$ are occupied.}
\end{figure}

We assume the first $n$ bands completely filled. For insulators, the calculations are more involved since the perturbing potential $w(x)$ may generate bound states in the insulating gap ($\equiv$ the gap above the $n$-th band). There can be a discrete or a continuum set of states inside the insulating gap. When the set is discrete, there are four distinct possibilities, as shown in Fig.~5. Let us analyze these cases first.

{\it No bound states in the insulating gap:} We can take ${\cal C}_1^{\prime}$ (with opposite orientation, see Fig.~2b) as the contour of integration in Eq.~(\ref{prima}). Mapping into the complex $\lambda$-plane and using the fundamental property of the Bloch functions, gives
\begin{equation}
    \Delta n(x)=\lambda_n^{2m}\frac{2i}{b}\int\limits_{{\cal C}_1}R_n^+(\lambda)\psi
    _{n,\lambda}(y)^2 \left( \frac{\lambda}{\lambda
    _{n}}\right)^{2m-1}\frac{d\lambda }{\lambda_n},
\end{equation}
with ${\cal C}_1$ shown in Fig.~2a and $y$ and $m$ defined in Eq.~(\ref{m}). The integrand diverges at $\lambda_n$ as $(\lambda -\lambda _{n})^{-1/2}$ but this singularity is integrable. Away from the branch point, the integrand is finite and $(\lambda /\lambda_{n})^{2m}$ becomes small as we follow the contour ${\cal C}_1$ towards $\lambda =0$. Thus, for large $m$, the main contribution to the integral comes from the region in the immediate vicinity of the branch point. Expanding the integrand near $\lambda_n$ and keeping the leading term,\cite {Olver97} we find
\begin{equation}
    \Delta n(x)\rightarrow R_n^+(\lambda_n)\frac{2}{b}\int\limits_{{\cal C}_1}
    \frac{i(\frac{\lambda}{\lambda_n})^{2m-1}}{\sqrt{\frac{\lambda}{\lambda_n}-1}}
    \frac{d\lambda}{\lambda_n}u_n(x)^2 e^{-2q_nx}
\end{equation}
The integral is equal to $-2B(2m,1/2)$ ($B$ = Beta function) and behaves asymptotically as $-\sqrt{2\pi /m}$. We conclude,
\begin{equation}\label{asymptotic4}
    \Delta n(x) \rightarrow -2R_n^+(\lambda_n)\left(
    \frac{2\pi}{xb}\right)^{1/2}u_n(x)^2 e^{-2q_nx}.
\end{equation}
Similarly,
\begin{equation}\label{asymptotic44}
    \Delta \epsilon(x) \rightarrow -2R_n^+(\lambda_n)\tilde{E}_n\left(
    \frac{2\pi}{xb}\right)^{1/2}u_n(x)^2 e^{-2q_nx}.
\end{equation}
An implementation of Eq.~(\ref{asymptotic4}) to the perturbed Kronig-Penney model described in Eq.~(\ref{vtotal}) is given in Fig.~6. Note again that the asymptotic regime starts from one or two lattice sites from the perturbation.

\begin{figure}
  \includegraphics[width=8.0cm]{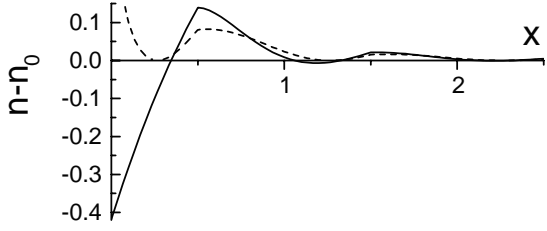}\\
  \caption{The exact (solid line) and asymptotic (dashed line) $\Delta n(x)$ for the perturbed Kronig-Penney model Eq. (\ref{vtotal}) [$v_0=-3$, $b=1$, $V_0=50$, first band completely filled].}
\end{figure}

To end the proof of nearsightedness, we need to show that $|R^+(\lambda_n)|$ cannot exceed an upper bound. Since the reflection coefficient is not evaluated on the unit circle, the inequality $|R^+(\lambda)|\leq 1$ is no longer guaranteed. However, if $h^<_\lambda$ and $h_{\lambda}$ denote the logarithmic derivatives at $x=0$ of $\psi
_\lambda^<(x)$ defined in Eq.~(\ref{psiless}), and of $\psi_{\lambda}(x)$, respectively, then
\begin{equation}\label{reflection}
    R^{+}(\lambda)=-\frac{h^<_\lambda -h_{1/\lambda}}{h^<_\lambda -h
    _{\lambda }}\frac{\psi_{1/\lambda}(0)}{\psi_\lambda(0)}.
\end{equation}
As in Ref.~\onlinecite{Kohn59}, we choose the phase of the Bloch functions such that $\psi_\lambda(0)=\psi_{1/\lambda}(0)$, so we can eliminate the last factor in the above expression.
For $E$ in the insulating gap, $\psi_\lambda^<(x)$, $\psi
_{\lambda}(x)$ and $\psi _{1/\lambda}(x)$ are real functions and
since
\begin{subequations}
\begin{equation}
    d
    h^<_\lambda/dE=-\psi_\lambda^<(0)^{-2}\int_{-\infty}^0\psi_\lambda^<(x)^2dx
\end{equation}
\begin{equation}
    d h_{1/\lambda}/dE=-\psi_{1/\lambda}(0)^{-2}\int_{-\infty}^0\psi
    _{1/\lambda}(x)^2dx
\end{equation}
\begin{equation}
    dh_{\lambda}/dE=\psi_\lambda(0)^{-2}\int_0^{\infty}\psi
    _{\lambda}(x)^2dx,
\end{equation}
\end{subequations}
it follows that $h_\lambda=-h_{1/\lambda}$ and $h^<_\lambda$ and $h_{1/\lambda}$ are decreasing functions of $E$. The typical behavior of $h_{\lambda}$ and $h_{1/\lambda}$ is shown in Fig.~7. Now, if there are no bound states in the gap, $h^<_\lambda$ and $h_{\lambda}$ cannot be equal for any $E$ in the gap. Then, since $h_{\lambda}$ increases while $h^<_\lambda$ decreases with $E$, $h^<_\lambda$ can take values only in the shaded area of Fig.~7, below 0. Consequently, the right side of Eq.~(\ref{reflection}) is smaller or equal to 1, i.e. 
\begin{equation}\label{upper}
     |R^{+}(\lambda)|\leq 1,
\end{equation}
remains valid when $E$ is in the insulating gap. 

We can then conclude that the amplitudes of $\Delta n(x)$ and $\Delta \epsilon(x)$, in the asymptotic limit, cannot exceed the upper bound
\begin{equation}\label{upper2}
    \overline{\Delta n}(x) \rightarrow 2\left(
    \frac{2\pi}{xb}\right)^{1/2}u_n(x)^2 e^{-2q_nx},
\end{equation}
and
\begin{equation}
    \overline{\Delta \epsilon}(x) \rightarrow \tilde{E}_n
    \overline{\Delta n}(x).
\end{equation}
This completes the proof of NEM for insulators and one sided perturbing potentials that do not generate
bound states in the insulating gap.

\begin{figure}
  \includegraphics[angle=90,width=7.0cm]{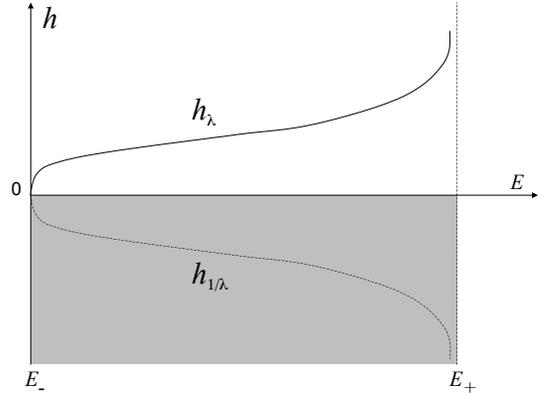}\\
  \caption{Typical behavior of $h_{\lambda}$ (solid lines) and $h_{1/\lambda}$ dashed lines. $E_{\pm}$ denote the upper/lower edge of the insulating gap. $h^<_E$ can take values only in the shade areas.}
\end{figure}

{\it Bound states in the insulating gap:} We show in Appendix B that if
\begin{equation}
\int w(x) \psi_{\pm}(x)^{2}dx \gtrless 0,
\end{equation}
where $\psi_{\pm}(x)$ denotes the Bloch function at the upper/lower edge of an energy band, then $w$ generates bound states above/below this band, even for infinitely small coupling constants. Thus, the
presence of bound states in the gaps is not a rare occurrence in one dimension.

The asymptotic forms of $\Delta n(x)$ and $\Delta \epsilon(x)$ depend on how the bound states in the insulating gap are occupied. When all bound states below the branch point $\tilde{E}_n$ are
occupied and the ones above $\tilde{E}_n$ are unoccupied, i.e. the situation illustrated in Fig.~5b, the
asymptotic behavior, Eqs.~(\ref{asymptotic4}) and (\ref{asymptotic44}), of $\Delta n(x)$ and $\Delta \epsilon(x)$ remains unchanged.

Consider now that there are unoccupied bound states below $\tilde{E}_n$, as illustrated in Fig.~5c. Let $\varphi_0$, of energy $E_0$, be such a state. For $x>0$, $\varphi_0$ is equal, up to a factorization constant, to the exponentially decaying Bloch function of energy $E_0$ ($E_0=E_n(\lambda_0)$):
\begin{equation}
    \varphi_0(x)=\left[\frac{(1-\lambda_0^2)\Lambda}{\int_0^b|\psi
    _{n,\lambda_0}(x)|^2dx}\right]^{1/2}\psi_{n,\lambda_0}(x),
\end{equation}
where
\begin{equation}
    \Lambda \equiv \int_0^{\infty }|\varphi_0(x)|^2dx \ \
    (\Lambda \leq 1).
\end{equation}
We have $|\lambda_0|=e^{-q_0b}$, with $q_0$ strictly larger than zero, and $\psi_{\lambda_0}(x)=e^{-q_0x}u_0(x)$, with $u_0(x+b)=(-1)^n u_0(x)$. Since $q_0$ decreases as $E_0$ moves away from $\tilde{E}_n$, the first unoccupied state will have the slowest exponential decay, among all unoccupied states below $\tilde{E}_n$. Thus, when the contribution of these states is subtracted from Eq.~(\ref{asymptotic4}), one finds that the asymptotic form of $\Delta n(x)$ is determined by the first unoccupied bound state:
\begin{equation}\label{asymptotic5}
    \Delta n(x)\rightarrow-\frac{2(1-\lambda_0^2)\Lambda}{\int_0^b|\psi
    _{n,\lambda_0}(x)|^2dx}u_0(x)^2e^{-2q_0x},
\end{equation}
with the index $0$ referring to the first unoccupied bound state in the insulating gap. Similarly,
\begin{equation}
    \Delta \epsilon(x)\rightarrow-\frac{2(1-\lambda_0^2)E_0\Lambda}{\int_0^b|\psi
    _{n,\lambda_0}(x)|^2dx}u_0(x)^2e^{-2q_0x}.
\end{equation}
Since $\Lambda \leq 1$, the amplitudes of $\Delta n(x)$ and $\Delta \epsilon (x)$ cannot exceed, in the asymptotic limit, the upper bounds
\begin{equation}\label{upper3}
    \overline{\Delta n}(x)\rightarrow \frac{2(1-\lambda_0^2)}{\int_0^b|\psi
    _{n,\lambda_0}(x)|^2dx}u_0(x)^2e^{-2q_0x},
\end{equation}
and
\begin{equation}
    \overline{\Delta \epsilon}(x)\rightarrow E_0\overline{\Delta n}(x).
\end{equation}

The results remain the same if, instead of unoccupied bound states below $\tilde{E}_n$, there are occupied bound states above $\tilde{E}_n$, as in Fig.~5d. In this case, the index 0 will refer to the last occupied bound state.

\begin{figure}
  \includegraphics[width=8.6cm]{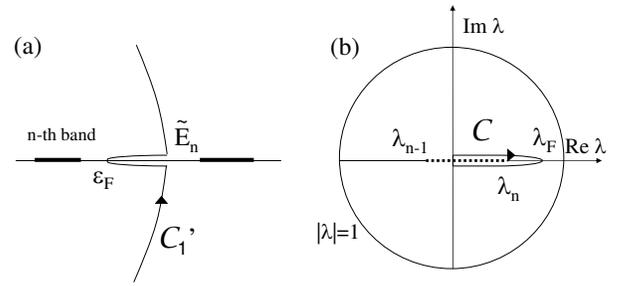}\\
  \caption{a) The contour of integration in the complex $E$-plane  (${\cal C}_1^{\prime}$ was introduced in Fig.~2). b) The same contour
  in the complex $\lambda$-plane.}
\end{figure}

{\it Continuum states in the insulating gap:} We consider the case when $w(x)$ fills the entire insulating gap with continuum spectrum, such as when the insulator is in contact with an infinite metal. In this case, $R_n^+(\lambda)$ has a branch cut on the real axis. The states are considered occupied up to a Fermi energy, $\epsilon_F$, which is in the insulating gap of the unperturbed insulator. We consider only the generic case when $\epsilon_F\neq \tilde{E}_n$ and define $\lambda_F$ by $\epsilon _F=E_{n}(\lambda_F)$. This $\lambda_F$ is located strictly inside the unit circle, as opposed to the case of metals. We also define $q_F>0$ so that $|\lambda_F|=e^{-q_Fb}$.

In Eq.~(\ref{prima}), we consider the contour of integration  shown in Fig.~8a. Mapping into the complex $\lambda$-plane and using again the fundamental property of the Bloch functions,
\begin{equation}
    \Delta n(x) =\lambda_F^{2m} \frac{2i}{b}\int\limits_{{\cal C}}R_n^+(\lambda) \psi
    _{n,\lambda}(y)^2\left (\frac{\lambda}{\lambda _F}\right)
    ^{2m-1}\frac{d\lambda }{\lambda_F}.
\end{equation}
With the new variable $q$ defined by $\lambda/\lambda_F=e^{-qb}$, we can write
\begin{equation}\label{generic}
    \Delta n(x) =-4\lambda_F^{2m}\text{Im}\int\limits_{q>0}R_n^+(
    \lambda^+) \psi_{n,\lambda}(y)^2 e^{-2mqb}dq,
\end{equation}
where $\lambda^+\equiv\lambda+i0^+$. The asymptotic behavior can be extracted from a simple integration by parts:
\begin{equation}
    \Delta n(x) \rightarrow -2\text{Im}[R_{n}^{+}(\lambda_F^+)]
    \frac{\psi_{n,\lambda_F}(x)^2}{x}.
\end{equation}
By writing $\psi_{n,\lambda_F}(x)$ as $u_{F}(x)e^{-q_F x}$, with $u_{F}(x+b)=(-1)^n u_{F}(x)$, we can conclude:
\begin{equation}\label{asymov}
    \Delta n(x) \rightarrow -2\text{Im}[R_{n}^{+}(\lambda_F^+)]
    u_{F}(x)^2 \frac{e^{-2q_F x}}{x}.
\end{equation}
Similarly,
\begin{equation}
    \Delta \epsilon(x) \rightarrow -2\text{Im}[R_{n}^{+}(\lambda_F^+)]
    \epsilon_F u_{F}(x)^2 \frac{e^{-2q_F x}}{x}.
\end{equation}
An implementation of Eq.~(\ref{asymov}) to the perturbed Kronig-Penney model Eq.~(\ref{vtotal}) is shown in Fig.~9. Notice again that the asymptotic regime starts from two lattice constants away from the perturbation.

\begin{figure}
  \includegraphics[width=8.0cm]{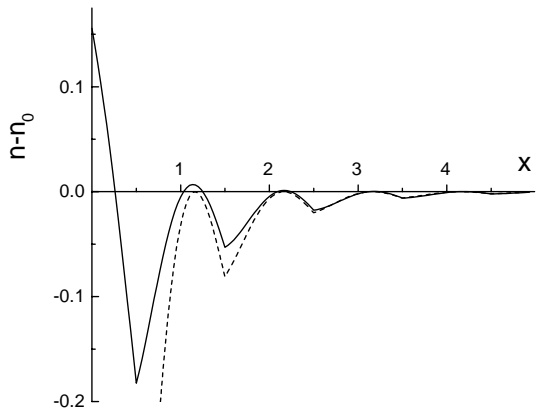}\\
  \caption{The exact (solid) and asymptotic (dashed) $\Delta n(x)$ for the model of Eq.~(\ref{vtotal})
  [$V_{0}=-5$, $v_0=-2$, $b=1$ and $\epsilon_F=4$].}
\end{figure}

To end the proof of NEM, we need to give an upper bound on the amplitudes of $\Delta n(x)$ and $\Delta\epsilon(x)$. The imaginary part of $R_{n}^{+}(\lambda_F^+)$ is proportional to the local density of states, $g(E,x)$, at $E=\epsilon_F$ and $x=0$. Indeed, Eq.~(\ref{Fundamental}) provides the following identity:
\begin{equation}
    \text{Im}[R_{n}^{+}(\lambda_F^+)]
    \psi_{n,\lambda_F}(x)^2=\frac{1}{2\pi}\frac{d\epsilon_F}{dq_F}
    \text{Im}[G_{\epsilon_F+i0}(x,x)],
\end{equation}
leading to
\begin{equation}\label{ImR}
    \text{Im}[R_{n}^{+}(\lambda_F^+)]=\frac{d\epsilon_F/dq_F}
    {2\psi_{n,\lambda_F}(0)^2}g(\epsilon_F,0).
\end{equation}
Note that the coefficient in front of $g(\epsilon_F,0)$ is determined by the unperturbed system. If we limit ourselves to the generic case of $w^\prime$s that generate finite densities of states at $\epsilon_F$, then NEM follows from Eqs.~(\ref{asymov}) and (\ref{ImR}). For practical applications, we consider this argument sufficient. However, to achieve a full proof of NEM, we need to consider also the cases when $g(E,0)$ diverges (or becomes extremely large) as $E\rightarrow \epsilon_F$. For this special cases, the asymptotic form of Eq.~(\ref{generic}) cannot be extracted from a simple integration by parts and $\Delta n(x)$ is no longer given by Eq.~(\ref{asymov});
its specific functional form will depend on the type of singularity of $g(E,0)$. This special situations will not be discussed here.

\subsection{Two Sided Perturbations}

We consider here the case when the point $x$, where we evaluate $\Delta n(x)$ and $\Delta \epsilon(x)$, has perturbing potentials $w_L$ to the left and $w_R$ to the right, as schematically shown in
Fig.~10. For one sided perturbations, $\Delta n(x)$ decays as $x$ moves further and further away from the perturbation and, because of this simple picture, NEM is intuitive and simply to grasp. When left and right perturbing potentials are present, this simple picture is gone: there will be interference terms in $\Delta n(x)$, whose amplitude remain constant in the region between the two perturbing potentials. In addition, $w_L+w_R$ can induce a strong, qualitative change of the system, namely, the energy bands may become quantized. In spite all of these, we will show the following: for metals, the interference terms are not negligible, but $\Delta n(x)$ still remains bounded. For insulators, the interference terms are exponentially small and $\Delta n(x)$ is given by the simple superposition of the left and right density changes. Similar conclusions hold for $\Delta \epsilon (x)$.

For convenience, we fix the origin in the middle of the interval that separates the two perturbing potentials and consider the distance, $\textsf{R}$, from the origin to the right/left edge of $w_{L/R}$ to be an integer of $b$, $\textsf{R}=Nb$. We are interested in the behavior, for $x$ near the origin, of $\Delta n(x)$ and $\Delta \epsilon(x)$ when $\textsf{R}\rightarrow \infty$. We follow our general strategy and derive first an expression for the Green's function on the
interval $[-\textsf{R},\textsf{R}]$. We look again for the solutions
of the Schrodinger equation,
\begin{equation}
    [-d^2/dx^2+v(x)+w_L(x)+w_R(x)]\psi(x)=E\psi(x),
\end{equation}
which decay at $\mp \infty$. On the intervals from $-\infty$ to the
left edge of $w_L$ and from the right edge of $w_R$ to $+\infty$,
these solutions can be expressed as in Eqs.~(\ref{psiless}) and
(\ref{psimore}), in terms of the total (corresponding to $w_L+w_R$)
transmission and reflection coefficients. Then, one can use again the reflection and transmission coefficients to continue these solutions inside the interval $-\textsf{R}<x<\textsf{R}$. On this interval, they take the following form:
\begin{equation}\label{psi1}
    \psi^>_\lambda(x)=\frac{T(\lambda)}{\tilde{T}_R(\lambda)}
    [\psi_\lambda(x)+\tilde{R}_R^-(\lambda)\psi_{1/\lambda}(x)]
\end{equation}
and
\begin{equation}\label{psi2}
    \psi^<_\lambda(x)=\frac{T(\lambda)}{\tilde{T}_L(\lambda)}
    [\psi_{1/\lambda}(x)+\tilde{R}_L^+(\lambda)\psi_{\lambda}(x)],
\end{equation}
where $\tilde{T}_{L,R}(\lambda)$ and $\tilde{R}^\pm_{L,R}(\lambda)$
are the transmission and reflection coefficients of the left/right
potentials, and $T(\lambda)$ is the total transmission coefficient,
\begin{equation}
    T(\lambda)=\frac{\tilde{T}_L(\lambda)\tilde{T}_R(\lambda)}
    {1-\tilde{R}_L^+(\lambda)\tilde{R}_R^+(\lambda)}.
\end{equation}
All these coefficients depend on $\textsf{R}$: If
$R^\pm_{L,R}(\lambda)$ denote the reflection coefficients when the
right/left edge of $w_{L/R}$ is at $x=0$ (thus
$R^\pm_{L,R}(\lambda)$ are independent of $\textsf{R}$), then
Eq.~(\ref{reflection}) gives:
\begin{equation}
    \tilde{R}^\pm_{L,R}(\lambda)=\lambda^{2N}R^\pm_{L,R}(\lambda).
\end{equation}
The Wronskian of the two independent solutions is the same as given
in Eq.~(\ref{Wronskian}). From the two independent solutions,
Eqs.~(\ref{psi1}) and (\ref{psi2}), and their Wronskian, we derive,
for $-\textsf{R}<x<\textsf{R}$, the following identity:
\begin{eqnarray}\label{DiffG}
    & &[G_E(x,x)-G_E^0(x,x)]\frac{dE}{\pi i}\\
    &=&\frac{2i\lambda^{2N}R_L^+(\lambda)}{1-\lambda^{4N}R_L^+(\lambda)R_R^-(\lambda)}
    \psi_\lambda(x)^2 \frac{d\lambda}{b \lambda}\nonumber \\
    &+&\frac{2i\lambda^{2N}R_R^-(\lambda)}{1-\lambda^{4N}R_L^+(\lambda)R_R^-(\lambda)}
    \psi_{1/\lambda}(x)^2 \frac{d\lambda}{b\lambda}\nonumber \\
    &+&\frac{4i\lambda^{4N}R_L^+(\lambda)R_R^-(\lambda)}{1-\lambda^{4N}R_L^+(\lambda)R_R^-(\lambda)}
    \psi_\lambda(x)\psi_{1/\lambda}(x)\frac{d\lambda}{b\lambda}.\nonumber
\end{eqnarray}
We now can use Eqs.~(\ref{prima}) and (\ref{duo}) to find $\Delta
n(x)$ and $\Delta \epsilon (x)$, for $x$ near the origin. By using the fundamental property of the Bloch functions, we can understand the behavior of each term in the above identity. For $\lambda$ not on the unit circle, as it is the case in our integrals, the first term decay exponentially as $x$ moves to the right; the second term decay exponentially as $x$ moves to the left, but the third term is periodic, with period $b$. Fortunately, the amplitude of this term becomes smaller and smaller as the two perturbing potentials are moved apart. 

\begin{figure}
  \includegraphics[width=8.6cm]{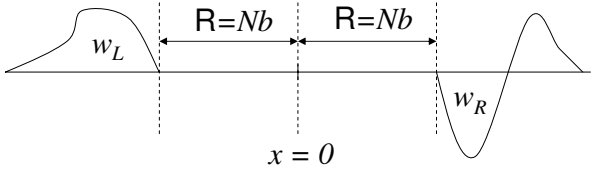}\\
  \caption{The case when the point $x$ (near the origin), where we evaluate $\Delta n(x)$ and
  $\Delta \epsilon (x)$, has perturbing potentials to the left and
  to the right.}
\end{figure}

From Eq.~(\ref{DiffG}), one can easily obtain the energy spectrum of the perturbed system. So, let us  discuss first how the simultaneous presence of $w_{L/R}$ affects the energy spectrum. We are interested in the last occupied band (indexed by $n$), so, from now on, $\lambda$ is considered on the $n$-th Riemann sheet. The energies of the discrete  states are given by the poles of the Green's function. From the identity Eq.~(\ref{DiffG}), one can see that these poles correspond to those $\lambda$ satisfying the equation
\begin{equation}\label{poles}
    R_L^+(\lambda)R_R^-(\lambda)=1/\lambda^{4N}.
\end{equation}
Since the discrete state energies are real, the solutions of Eq.~(\ref{poles}) are always located on the unit circle or on the real axis, away from the branch cuts. Inside the unit disk we have $|\lambda|<1$; consequently, $|1/\lambda^{4N}|$ becomes very large in the limit $\textsf{R}\rightarrow \infty$. Thus, if there are solutions of Eq.~(\ref{poles}) inside the unit disk, they must be located very close to the poles of either $R^+_L(\lambda)$ or $R^-_R(\lambda)$. In other words, these solutions are perturbations of the bound states generated by $w_{L}$ (or $w_R$) alone, already discussed in the previous subsection. Poles on the unit circle exist if and only if both $|R^\pm_{L,R}(\lambda)|$ are equal to 1. In this case, the energy band degenerates into discrete energy spectrum. Because the left hand side of Eq.~(\ref{poles}) is slowly varying compared to the right hand side, one can get the qualitative picture by setting the left hand side constant. If  the amplitudes of $R^\pm_{L,R}(\lambda)$ are equal to 1 on the whole unit circle, then Eq.~(\ref{poles}) has $4N$ solutions, $\lambda_k$, distributed on the unit circle; the spacing between two consecutive solutions is $2\pi/4N+O(1/N^2)$. This is the picture in the $\lambda$-plane. In the $E$-plane, the discrete energies are given by $E(\lambda_k)$. The spacing between two consecutive energies is $2\pi \partial_\lambda E(\lambda_k)/4N+O(1/N^2)$.

\subsubsection{Metals}

The effects of band quantization will be the strongest for metals, since the Fermi energy lies inside the band. We calculate the density change by integrating Eq.~(\ref{DiffG}) along the contour of integration shown in Fig.~3a and map the integral into the complex $\lambda$-plane. The asymptotic behavior of $\Delta n(x)$, for large $\textsf{R}$, is determined by the behavior of the integrand near $\lambda_F$ and $1/\lambda_F$ (see Fig.~3b). In the immediate vicinity of these points, we can replace the slowly varying functions in the right hand side of Eq.~(\ref{DiffG}) [the reflection coefficients and the Bloch functions] with their value at $\lambda_F$ and $1/\lambda_F$. The integral then can be explicitly calculated and the result is:
\begin{eqnarray}\label{LRDn}
    \Delta n(x,\textsf{R}) &\rightarrow&
    \frac{2}{\textsf{R}}\text{Im}\frac{\tanh^{-1}[\lambda_F^{2N}(R_L^+R_R^-)^{1/2}]}
    {(R_L^+R_R^-)^{1/2}}  \\
    &\times& [R_L^+\psi_{\lambda_F}(x)^2+R_R^-\psi_{1/\lambda_F}(x)^2]
    \nonumber \\
    &+&\frac{2}{\textsf{R}}\text{Im}[\ln(1-\lambda_F^{4N}R_L^+R_R^-)]
    |\psi_{\lambda_F}(x)|^2, \nonumber
\end{eqnarray}
with the reflection coefficients evaluated at $\lambda_F$. 

To prove NEM, we need to find an upper bound on the above expression. A simple analysis reveals that the largest density changes occur when $|R^{\pm}_{L,R}|=1$, i.e. when the band is quantized at the Fermi energy. In this case, we can rewrite Eq.~(\ref{LRDn}) as:
\begin{eqnarray}\label{asymptotic11}
    \Delta n(x,\textsf{R}) &\rightarrow&
    \frac{4}{\textsf{R}}\text{Im}\left[\tanh^{-1}[
    \lambda_F^{2N}(R_L^+R_R^-)^{1/2}]\right] \\
    &\times& \text{Re}[(R_L^+R_R^{-\ast})^{1/2}\psi_{\lambda_F}(x)^2]
    \nonumber \\
    &+&\frac{2}{\textsf{R}}\text{Im}[\ln(1-\lambda_F^{4N}R_L^+R_R^-)]
    |\psi_{\lambda_F}(x)|^2. \nonumber
\end{eqnarray}
The density change, as function of $\textsf{R}$, has discontinuities every time when $\lambda_F^{4N}R_L^+R_R^-=1$, i.e. when a discrete energy crosses the Fermi level. These discontinuities are finite:
since $|\text{Im}[\tanh^{-1}(z)]|\leq \pi/4$ and $|\text{Im}[\ln(1-z)]|\leq \pi/2$, for $|z|\leq1$, the amplitude of the asymptotic term of $\Delta n(x,\textsf{R})$ cannot exceed the upper bound 
\begin{equation}\label{asyfinal1}
    \overline{\Delta n}(x,\textsf{R})\rightarrow \frac{2\pi}{\textsf{R}}|\psi_{\lambda_F}(x)|^2,
\end{equation}
independent of $w_{L/R}$ potentials and of the position of the Fermi energy relative to the discretized energies. Similarly, the amplitude of $\Delta \epsilon (x,\textsf{R})$ cannot
exceed the upper bound
\begin{equation}
    \overline{\Delta \epsilon}(x,\textsf{R})\rightarrow
    \epsilon_F\overline{\Delta n}(x,\textsf{R}),
\end{equation}
and this completes our discussion of NEM for metals. 

These upper bounds are optimal, in the sense that there are $w_{L/R}$ potentials (the worst scenario) that generate a density and an energy density that are equal to these upper bounds. By comparing with the results of the previous Section, one can see that interference has non-trivial effects: these upper bonds are not simply the superposition of the left and right upper bounds.

\subsubsection{Insulators}
We consider first the situation when there are no bound states in the insulating gap. We can take ${\cal C}_1^{\prime}$ of Fig.~2 as the contour of integration in Eq~(\ref{prima}), which is mapped into
${\cal C}_1$ in the complex $\lambda$-plane. For any $\lambda$ on
this curve, the denominators in the right side of Eq.~(\ref{DiffG})
goes exponentially to 1 as $\textsf{R}\rightarrow \infty$. Consequently, in this limit, the
structure of the integrand becomes completely analogous with the one
studied in the previous subsection. The asymptotic behavior can be
extracted as previously and the result is
\begin{eqnarray}
    \Delta n(x,\textsf{R}) \rightarrow
    -2R_L^+(\lambda_n)\left(\frac{2\pi}{b\textsf{R}}\right)^{1/2}u_n(x)^2 e^{-2q_n
    \textsf{R}} \nonumber \\
    -2R_R^-(\lambda_n)\left(\frac{2\pi}{b\textsf{R}}\right)^{1/2}u^{\prime}_n(x)^2 e^{-2q_n
    \textsf{R}} \nonumber \\
    -2R_L^+(\lambda_n)R_R^-(\lambda_n)\left(\frac{\pi}{b\textsf{R}}\right)^{1/2}u_n(x)u^{\prime}_n(x) e^{-4q_n
    \textsf{R}}.
\end{eqnarray}
Thus, in the limit of large $\textsf{R}$, $\Delta n(x,\textsf{R})$
is just the sum of the independent density changes due to the left
and right potentials, plus an exponentially small correction. From
the previous subsection, we can conclude that the amplitude of
$\Delta n(x,\textsf{R})$ cannot exceed, for large $\textsf{R}$, the
upper bound
\begin{equation}\label{asyfinal2}
    \overline{\Delta n}(x,\textsf{R}) \rightarrow
    2\left(\frac{2\pi}{b\textsf{R}}\right)^{1/2}[u_n(x)^2+u_n^\prime(x)^2] e^{-2q_n
    \textsf{R}}.
\end{equation}
Similarly,
\begin{equation}
    \overline{\Delta \epsilon}(x,\textsf{R}) \rightarrow
    \tilde{E}_n\overline{\Delta n}(x,\textsf{R}).
\end{equation}
For the case when there are bound or continuum states in the insulating gap, the conclusion is the same: for large $\textsf{R}$,, the density change near the origin is the sum of the independent changes induced by the left and right potentials. Upper bounds on $\Delta n(x,\textsf{R})$ can be trivially derived from the previous subsection.

\section{The Nearsightedness Range}

The nearsightedness range $\textsf{\textsf{R}}(x,\Delta n)$ was
introduced as the range beyond which any perturbation, no matter how
large, induces a density change at $x$ less than the given $\Delta
n$. The asymptotic $\textsf{\textsf{R}}(x,\Delta n)$, in the limit
$\Delta n\rightarrow 0$, can now be easily calculated from the upper
bounds on $\Delta n(x)$, derived in this paper. Since the periodic
systems are macroscopically homogeneous, the asymptotic
$\textsf{\textsf{R}}(x,\Delta n)$ will be independent of $x$.

When solving for $\textsf{R}$ in $\overline{\Delta
n}(x,\textsf{R})=\Delta n$, we first average $\overline{\Delta
n}(x,\textsf{R})$ over one unit cell. For metals,
Eq.~(\ref{asyfinal1}) leads to the following asymptotic expression:
\begin{equation}\label{near2}
    \textsf{R}(\Delta n)\rightarrow1/\Delta n.
\end{equation}
Such universal behavior is characteristic only to 1 dimension; in
higher dimensions, the nearsightedness range will depend on the
average particle density.\cite{ProdanKohn}

For insulators and $w's$ that generate no bound states in the
insulating gap, Eq.~(\ref{asyfinal2}) leads to
\begin{equation}\label{near1}
    \textsf{R}(\Delta n)\rightarrow\frac{1}{2q_n}\ln \frac{\tilde{n}}{\Delta n},
\end{equation}
where
\begin{equation}\label{tilden}
    \tilde{n}=\frac{4\sqrt{2 \pi q_n}}{b}\int_{0}^{b}[u_n(x)^2+u_n^\prime(x)^2] \ dx.
\end{equation}
In the small gap and tight binding limits, $\tilde{n}$ is completely
determined by the exponential decay constant $q_n$,
$\tilde{n}\rightarrow 4q_n\sqrt{2/\pi}$ and $\tilde{n}\rightarrow
4\sqrt{q_n/\pi b}$, respectively.

It is important to notice that the nearsightedness range does not depend on the details of the underlying potential $v(x)$, but on some simple parameters that can be defined also for non periodic potentials. For example, $q_n$ can be identified with the exponential decay constant of the density matrix.

\section{Discussion}

The above analysis provides a quantitative analysis of the nearsightedness of electronic matter for non-interacting fermions, moving in one dimension under the action of periodic potentials. Although the simplest case possible, it allowed us to understand the different mechanisms behind NEM. Although we cannot point to one simple and general physical explanation of NEM,  it is now clear that NEM is due to a destructive interference not of the wave amplitudes but of density amplitudes $n_j$ associated with the single particle eigenstates $\psi_j $. The asymptotic behavior of $\Delta n(x)$ was found to be determined by the reflection coefficient. For specific cases, the amplitude of $\Delta n(x)$ cannot exceed an upper bound simply because, when evaluated at allowed energies, the reflection coefficients are always smaller than 1. More general, and now including 2 and 3D, one will find that asymptotic behavior of $\Delta n(x)$ is determined by certain elements of the scattering matrix and, for specific cases, the unitarity of the scattering matrix imposes certain upper  bounds. The situation is, however, more complicated when bound states appear in the insulating gap or when the bands become quantized. 

We have introduced a new concept, the nearsightedness range, $\textsf{R}(x,\Delta n)$, which is well defined only because there is this upper bound on $\Delta n(x)$. $\textsf{R}(x,\Delta n)$ is a characteristic of the unperturbed system and gives a simple and effective measure of nearsightedness. For periodic metals, we found $\textsf{R}(x,\Delta n)$ to have, in the asymptotic limit $\Delta n\rightarrow 0$, a universal expression, namely $1/\Delta n$. For insulators, $\textsf{R}(x,\Delta n)$ is strongly dependent on the band structure but has a weak, logarithmic dependence on $\Delta n$.

Although the estimates given in this paper can be applied only to 1D systems, we think we gain some knowledge that can be useful for more general situations. We are convinced that NEM exists in dimensions higher than 1, where it can be quantified in a similar way. In particular, we believe that a complete theoretical analysis and optimization of the O(N) divide and conquer algorithm is possible in all dimensions. Preliminary results in this direction have been already given in Ref.~\onlinecite{ProdanKohn}. The one dimension analysis proved to be extremely useful by providing a viable strategy and some understanding of the effects of the bound states in the in insulating gap and of the band quantization on NEM.

\begin{acknowledgments}
I would like to thank Prof. Walter Kohn, who suggested and
supervised this project. This work was completed when the author was visiting the Physics Department at the University of California at Santa Barbara and was supported by Grants No.
NSF-DMR03-13980, NSF-DMR04-27188 and DOE-DE-FG02-04ER46130.
\end{acknowledgments}

\appendix

\section{}

We estimate here the exponential decay rate $q_n$, related to the
branch point by $|\lambda_{n}| =e^{-q_{n}b}$. According to
Ref.~\onlinecite{Kohn59}:
\begin{equation}
q_n=\frac{1}{b}\ln [|\mu_n|+\sqrt{\mu_n^2-1}],
\end{equation}
where $\mu_{n}$ is the Krammers function evaluated at the branch
point $\tilde{E}_n$, defined by $d\mu/dE|_{E=\tilde{E}_n}=0$.

For small gaps, the behavior of $\mu(E)$ inside the entire gap is
well approximated by a quadratic function of $E$:
\begin{equation}\label{mu}
    \mu(E)\simeq (-1)^n\left[1-\frac{m_n^*b^2(E-E_n^+)
    (E-E_{n+1}^-)}{2(E_{n+1}^--E_n^+)}\right],
\end{equation}
where $E_n^\pm$ is the upper/lower edge of the $n$-th band, and
$m_n^*$ is the effective mass at the upper edge of the $n$-th band.
Since $\mu_n\gtrsim 1$, $q_n\simeq \frac{1}{b}\sqrt{2(|\mu_n|-1)}$,
which, together with Eq.~(\ref{mu}), lead to
\begin{equation}\label{q}
    q_n=\frac{1}{2}\sqrt{m_n^* (E_{n+1}^--E_n^+)}.
\end{equation}

We consider now a periodic potential $\sum_l V_a(x-lb)$, where
$V_a(x)$ vanishes for $|x|>c$ and has atomic levels $E_n\equiv
-k_n^2$, $n=1,\ldots$. In the limit $b\rightarrow \infty$, we show
that
\begin{equation}
q_{n}\simeq \frac{1}{b}\ln \frac{8\sqrt{-E_{n}}}{ebW_{n}},
\label{tightbiding}
\end{equation}
where $W_n$ is the width of the $n$-th energy band. For $x$ in
$[-b/2,b/2]$ and $|x|>c$, the solutions of the Schrodinger equation
at an energy $E=-k^{2}$ are of the general form
\begin{equation}
    \psi (x)=\left\{
    \begin{array}{c}
    a_-(k) e^{-kx}+b_-(k)e^{kx}, \ x<-c \\
    a_+(k) e^{-kx}+b_+(k) e^{kx}, \ x>c,
    \end{array}
\right.
\end{equation}
with
\begin{equation}
    \left(
    \begin{array}{c}
    a_+(k) \\
    b_+(k)
    \end{array}
    \right) =\hat{T}(k) \left(
    \begin{array}{c}
    a_-(k) \\
    b_-(k)
    \end{array}
    \right),
\end{equation}
$\hat{T}(k)$ being the transfer matrix of the potential $V_a$. The
energy levels of $V_a$ correspond to the zeroes of $T_{22}(k)$,
already denoted by $k_n$. The Kramers function is given by
\begin{equation}
    \mu (k) =\frac{1}{2}\left[ T_{11}(k) e^{-kb}+T_{22}(k)e^{kb}\right].
    \label{Kramer}
\end{equation}
We estimate first the bandwidths. We look for the solutions of
$\mu(k) =\pm 1$, which give the band edges. For $b$ large, the
solutions of this equation must be located very close to the zeros
of $T_{22}(k)$ since, otherwise, the second term in
Eq.~(\ref{Kramer}) becomes very large. We can then linearize,
$T_{22}(k) \simeq (k-k_n)T_{22}^{\prime }(k_n)$ and neglect the
exponentially small term in Eq.~(\ref{Kramer}), in which case the
equation $\mu(k) =\pm 1$ can be trivially solved, leading to
\begin{equation}\label{Bandwidth}
    W_n=\frac{8k_ne^{-k_n b}}{| T_{22}^\prime(k_n)| }.
\end{equation}
We now calculate $\tilde{E}_{n}=-\tilde{k}_{n}^{2}$, defined
by\cite{Kohn59}
\begin{equation}
    \left( \frac{d\mu }{dk}\right) _{k=\tilde{k}_{n}}=0\Leftrightarrow
    T_{22}^{\prime }(\tilde{k}_n) \simeq -bT_{22}(\tilde{k}_n).
\end{equation}
For $b$ large, the solutions of the above equation must also be
close to the zeroes of $T_{22}(k)$. Linearizing $T_{22}(k)$, we find
$\tilde{k}_n=k_n-1/b$ and the Kramers function evaluated at
$\tilde{k}_n$ is
\begin{equation}
    \mu_n=\frac{-T_{22}^{\prime }(k_n) e^{k_{n}b}}{2eb}=\frac{4k_{n}}{ebW_{n}}.
\end{equation}
Since $\mu_n\gg 1$, $q_n\simeq\frac{1}{b}\ln[2\mu_n]$ and
Eq.~(\ref{tightbiding}) follows.

\section{}

Let $w(x)$ be a perturbing potential of finite support and such that
\begin{equation}\label{condition1}
    \int w(x) \psi_{+}(x)^{2}dx > 0
\end{equation}
or
\begin{equation}\label{condition2}
    \int w(x) \psi_{-}(x)^{2}dx < 0,
\end{equation}
where $\psi_{\pm}(x)$ is the Bloch function at the upper/lower edge
of an energy band. We show here that, even for infinitely small
coupling constants, such potential will pull bound states out from
the band.

Let $H_0$ denote the periodic Hamiltonian and $H\equiv H_0+\gamma
w$. It is known that $H$ has a bound state at some energy $E$ if and
only if the operator\cite{SimonTr}
\begin{equation}
    \hat{K}_{E}=\gamma w^{1/2}(E-H_0)^{-1}|w|^{1/2}
\end{equation}
has an eigenvalue equal to 1.\cite{SimonTr} Here,
$w^{1/2}=w/|w|^{1/2}$. We show that, for any given energy $E$
below/above the band, $\hat{K}_E$ has an eigenvalue equal to 1 for
some positive $\gamma$, which decreases to zero as $E$ approaches
the edges of the band, provided the condition
Eq.~(\ref{condition1})/(\ref{condition2}) is satisfied.

If $n$ is odd, the lower edge of the band corresponds to
$\lambda=1$. We take an energy $E$ below such band and let
$\lambda$, which is real and less than 1, be its corresponding
$\lambda$-parameter. Eq.~(\ref{GreenF}) gives
\begin{equation}
    \hat{K}_E(x,x^{\prime}) =-\frac{2\pi\gamma}{b}w(x)^{1/2}\frac{\psi_{1/\lambda}(x_<)\psi
    _{\lambda }(x_>)}{\lambda dE/d\lambda}
    |w(x^{\prime})|^{1/2},
\end{equation}
and we notice that $dE/d\lambda\propto \lambda-1$, for
$\lambda\rightarrow 1$, i.e. the kernel $\hat{K}_E(x,x^{\prime})$
diverges at $\lambda=1$. We can separate the diverging part by
expanding
\begin{eqnarray}
    \psi_{\lambda ^{-1}}(x_<)
    \psi_{\lambda}(x_>)
    &=&\psi_-(x)\psi_-(x^{\prime}) \nonumber \\
    &+&(\lambda
    -1)W_{\lambda}(x,x^{\prime}).
\end{eqnarray}
This provides the following decomposition,
\begin{equation}\label{kappa}
    \hat{K}_{E}=\gamma\alpha(\lambda) |\varphi_1\rangle \langle
    \varphi_2|+\gamma \hat{A}(\lambda),
\end{equation}
where
\begin{equation}
    \alpha\equiv -\frac{2\pi}{b\lambda dE/d\lambda},
\end{equation}
\begin{equation}
    \hat{A}(\lambda)\equiv\frac{2\pi}{b}\frac{1-\lambda}{\lambda dE/d\lambda}w^{1/2}W_{\lambda}|w|^{1/2},
\end{equation}
and
\begin{equation}
    \left\{
    \begin{array}{l}
    \varphi_1(x)\equiv w(x)^{1/2}\psi_-(x) \\
    \varphi_2(x)\equiv|w(x)|^{1/2}\psi_-(x).
    \end{array}
    \right.
\end{equation}
The first term of Eq.~(\ref{kappa}) diverges while the second one is
analytic at $\lambda=1$. Now let $\Psi$ be given by
\begin{equation}
    \Psi=(\gamma\hat{A}(\lambda)-1)^{-1}\varphi_1,
\end{equation}
which is well defined for small $\gamma$. Then
\begin{equation}
    \hat{K}_{E}\Psi=\Psi+[1+\gamma\alpha(\lambda)\langle \varphi_2|(\gamma
    \hat{A}(\lambda)-1)^{-1}|\varphi_1\rangle]\varphi_1.
\end{equation}
In other words, $\hat{K}_{E}$ has an eigenvalue at +1, if
\begin{equation}\label{basics}
    1+\gamma\alpha(\lambda)\langle \varphi_2|(\gamma
    \hat{A}(\lambda)-1)^{-1}|\varphi_1\rangle=0.
\end{equation}
We can rewrite this equation as
\begin{equation}
    \gamma\langle \varphi_2|(\gamma
    \hat{A}(\lambda)-1)^{-1}|\varphi_1\rangle-\frac{b\lambda}{2\pi}\frac{dE}{d\lambda}=0.
\end{equation}
If we denote the left side with $F(\gamma,\lambda)$, then $F(0,1)=0$
and $\partial_\gamma F(0,1)=-\int w(x) \psi_{-}(x)^{2}dx>0$, i.e.
the conditions of the analytic implicit function theorem are
satisfied, which means that, for any $\lambda$ near $+1$, there is
always a solution $\gamma(\lambda)$ to the Eq.~(\ref{basics}).
Moreover,
\begin{equation}\label{gamma}
    \gamma(\lambda)=-\frac{b\lambda}{2\pi} \frac{dE}{d\lambda}\left[\int w(x)
    \psi_{-}(x)^{2}dx\right]^{-1}+\ldots,
\end{equation}
where the dots indicate terms of order $o[(1-\lambda)^2]$. $\gamma$
is real and positive, for $E$ below the band, and goes to zero as
$E$ approaches the band edge.

The other possible cases, $\lambda=-1$ and $n$ even, follow in the
same way.


\begin{references}
\bibitem{Kohn96}  W. Kohn, Phys. Rev. Lett. {\bf 76}, 3168 (1996).

\bibitem{ProdanKohn} E. Prodan and W. Kohn, PNAS {\bf 102}, 11635 (2005).

\bibitem{Friedel} J. Friedel, Phil. Mag. {\bf 43}, 153 (1952).

\bibitem{Rowland1} N. Bloemberg and T.J. Rowland, Acta Met. {\bf 1}, 731 (1953).

\bibitem{Rowland2} T.J. Rowland, Phys. Rev. {\bf 119}, 900 (1960).

\bibitem{Kohn60} W. Kohn, Phys. Rev. {\bf 119}, 912 (1960).

\bibitem{KohnHohenberg} P. Hohenberg and W. Kohn, Phys. Rev. {\bf 136}, B864 (1964).

\bibitem{KohnSham} W. Kohn and L.J. Sham, Phys. Rev. {\bf 140}, A1133 (1965).

\bibitem{Giullia} G. Galli, Phys. Stat. Sol. {\bf 217}, 231 (2000).

\bibitem{Yang91} W. Yang, Phys. Rev. Lett. {\bf 66}, 1438 (1991).

\bibitem{Goedecker} S. Goedecker, Rev. Mod. Phys. {\bf 71}, 1085 (1999).

\bibitem{Wu} S.Y Wu and C.S. Jayanthi, Phys. Reports {\bf 358}, 1 (2002).

\bibitem{Vashishta05} F. Shimojo, R.K. Kalia, A. Nakano and P. Vashishta, Comp. Phys. Comm. {\bf 167}, 151 (2005).

\bibitem{Kohn59}  W. Kohn, Phys. Rev. {\bf 115}, 809 (1959).

\bibitem{Prodan05} E. Prodan, cond-mat/0510446.

\bibitem{Allen} R.E. Allen, Phys. Rev. B {\bf 20}, 1454 (1979).

\bibitem{Kramers35}  H.A. Kramers, Physica {\bf 2}, 483 (1935).

\bibitem{Kronig}  R. de L. Kronig and W.G. Penney, Proc. Roy. Soc. {\bf A130}
(1931).

\bibitem{Olver97}  F.W.J. Olver, Asymptotic and Special Function: A.K.
Peters, Wellesley, MA (1997).

\bibitem{SimonTr}  B. Simon, Trace Ideals and Their Applications: Cambridge
Univ. Press, New York (1978).

\end{references}
\end{document}